\documentclass[reprint,amsmath,amssymb,aps,prd]{revtex4-1}
\usepackage{subfigure}
\usepackage{graphicx}
\usepackage{dcolumn}
\usepackage{float}
\usepackage{xcolor}
\usepackage{multirow}
\usepackage{enumerate}
\begin{document}
\title{Measurement of branching fractions of \boldmath$\psi(3686)\rightarrow\phi\eta^\prime, \phi f_1(1285)$ and \boldmath$\phi \eta(1405)$}
\author{M.~Ablikim$^{1}$, M.~N.~Achasov$^{10,d}$, P.~Adlarson$^{59}$, S. ~Ahmed$^{15}$, M.~Albrecht$^{4}$, M.~Alekseev$^{58A,58C}$, A.~Amoroso$^{58A,58C}$, F.~F.~An$^{1}$, Q.~An$^{55,43}$, Y.~Bai$^{42}$
% ...
\\
\vspace{0.2cm}
(BESIII Collaboration)\\
\vspace{0.2cm} {\it
$^{1}$ Institute of High Energy Physics, Beijing 100049, People's Republic of China\\
% ...
$^{4}$ Bochum Ruhr-University, D-44780 Bochum, Germany\\
% ...
$^{10}$ G.I. Budker Institute of Nuclear Physics SB RAS (BINP), Novosibirsk 630090, Russia\\
% ...
$^{15}$ Helmholtz Institute Mainz, Johann-Joachim-Becher-Weg 45, D-55099 Mainz, Germany\\
% ...
$^{42}$ Southeast University, Nanjing 211100, People's Republic of China\\
$^{43}$ State Key Laboratory of Particle Detection and Electronics, Beijing 100049, Hefei 230026, People's Republic of China\\
% ...
$^{55}$ University of Science and Technology of China, Hefei 230026, People's Republic of China\\
% ...
$^{58}$ (A)University of Turin, I-10125, Turin, Italy; (B)University of Eastern Piedmont, I-15121, Alessandria, Italy; (C)INFN, I-10125, Turin, Italy\\
$^{59}$ Uppsala University, Box 516, SE-75120 Uppsala, Sweden\\
% ...
$^{d}$ Also at the Novosibirsk State University, Novosibirsk, 630090, Russia\\
% ...
}}

\vspace{0.4cm}
\date{\today}

\begin{abstract}
Using a sample of $448.1\times10^6$ $\psi(3686)$ events collected with the BESIII detector, we perform a study of the decay $\psi(3686) \rightarrow \phi \pi^{+}\pi^{-}\eta$. The branching fraction of $\psi(3686)\rightarrow\phi\eta^\prime$ is determined to be $(1.51\pm0.16\pm0.12)\times 10^{-5}$, which is consistent with the previous measurement but with significantly improved precision.
The resonances $f_{1}(1285)$ and $\eta(1405)$ are clearly observed in the $\pi^{+}\pi^{-}\eta$ mass spectrum with statistical significances of $18\sigma$ and $9.7\sigma$, respectively. The corresponding product branching fractions are measured to be $\mathcal{B}(\psi(3686)\rightarrow\phi f_{1}(1285),f_{1}(1285)\rightarrow\pi^{+}\pi^{-}\eta) =(1.03\pm0.10\pm0.09)\times 10^{-5}$ and $\mathcal{B}(\psi(3686)\rightarrow\phi\eta(1405),\eta(1405)\rightarrow\pi^{+}\pi^{-}\eta) =(8.46\pm1.37\pm0.92)\times 10^{-6}$. These results are used to test the perturbative QCD ``$12\%$ rule''.
\end{abstract}

\maketitle

\section{\label{sec:introduction}Introduction}

In the quark model, if the vector meson nonet is ideally mixed, then the $\phi$ contains only strange quarks and the $\omega$ contains only up and down quarks. This assumption can be used as a `flavor filter' for the determination of the quark content of various resonant structures by observing their production in $J/\psi$ or $\psi(3686)$ decays in association with a $\phi$ or an $\omega$.

The BES experiment reported the measurement of the branching fraction (BF) of $\psi(3686)\rightarrow\phi\eta^\prime$ with $\eta^\prime\rightarrow\pi^+\pi^-\eta$ and $\eta^\prime\rightarrow\gamma\rho$, which was determined to be $(3.1\pm1.4\pm 0.7)\times10^{-5}$~\cite{phietap}.
Meanwhile, the decay $J/\psi\rightarrow\phi\pi^+\pi^-\eta$ has been used to investigate the $f_1(1285)$ and $\eta(1405)$ mesons~\cite{Jpsiprd}. Production of the $f_1(1285)$ state was clearly established, while the $\eta(1405)$ mode had a statistical significance of only 3.6$\sigma$.  A comparison with measurements in the $J/\psi\rightarrow\omega\eta\pi^+\pi^-$ channel~\cite{omegaetapipi} suggests that in $J/\psi$ decays the $\eta(1405)$ is preferentially produced in association with an $\omega$ rather than a $\phi$. This implies that the quark content of the $\eta(1405)$ is dominated by the lightest quarks, $u$ and $d$. Due to limited statistics, the higher mass region of the $\pi^+\pi^-\eta$ invariant mass spectrum has never been investigated in $\psi(3686)\rightarrow\phi\pi^+\pi^-\eta$ decays.

In addition, it is expected in perturbative QCD that both $J/\psi$ and $\psi(3686)$ decays to hadrons are via three gluons or a photon, which provides the relation~\cite{puzzle1,puzzle2}
\begin{equation}
Q_h=\frac{\mathcal{B}_{\psi(3686)\rightarrow h}}{\mathcal{B}_{J/\psi\rightarrow h}}=\frac{\mathcal{B}_{\psi(3686)\rightarrow e^+e^-}}{\mathcal{B}_{J/\psi\rightarrow e^+e^-}}\approx12\%.
\end{equation}
This relation is referred to as the ``$12\%$ rule'' and it is expected to hold to a reasonably good degree for both inclusive and exclusive decays. By studying the $\psi(3686)\rightarrow\phi\pi^+\pi^-\eta$ decay, we can check the ``$12\%$ rule'' in $\psi\rightarrow\phi\eta^\prime$, $\psi\rightarrow\phi f_1(1285)$ and $\psi\rightarrow\phi\eta(1405)$ decays.

A sample of $448.1\times10^6$ $\psi(3686)$ events collected with the BESIII detector~\cite{BESIII}, about 30 times larger than that of the BES experiment, offers a unique opportunity to improve the  precision of the BF of $\psi(3686)\rightarrow\phi\eta^\prime$ and to investigate the $\pi^+\pi^-\eta$ invariant mass spectrum above the $\eta^\prime$ mass.

In this paper, an improved measurement of the branching fraction $\mathcal{B}(\psi(3686)\rightarrow\phi\eta^\prime)$ is presented. We also report the observation of the $f_1(1285)$ and $\eta(1405)$ in the  $\pi^+\pi^-\eta$ mass spectrum and a measurement of the corresponding product branching fractions.

\section{\label{detector}Detector and Monte Carlo simulation}

BEPCII is a double-ring electron-positron collider with design peak luminosity of $10^{33}$ cm$^{-2}$s$^{-1}$ with a beam current of 0.93 A at $\sqrt{s}=3.773$ GeV. The cylindrical core of the BESIII detector consists of a helium-based main drift chamber (MDC), a plastic scintillator time-of-flight (TOF) system, a CsI(Tl) electromagnetic calorimeter (EMC), a superconducting solenoidal magnet providing a 1.0 T magnetic field, and a muon system made of resistive plate chambers in the iron flux return yoke of the magnet. The acceptances for charged particles and photons are 93\% and 92\% of $4\pi$, respectively. The charged particle momentum resolution is 0.5\% at $1~ \text{GeV}/c$, and the barrel (end cap) photon energy resolution is 2.5\% (5.0\%) at 1 GeV.

The optimization of the event selection and the estimation of the physics background are performed using Monte Carlo (MC) simulated samples. The {\sc geant4}-based~\cite{Agostinelli} simulation software {\sc boost}~\cite{BOSST} includes the geometry and material description of the BESIII detector, the detector response and digitization models, as well as a record of the detector running conditions and performance. To study the potential backgrounds, an inclusive MC sample of $506\times 10^6$ $\psi(3686)$ decays is generated, where the production of the $\psi(3686)$ resonances are simulated by the MC event generator {\sc kkmc}~\cite{Jadach2000,Jadach2001}. The known decay modes are generated by {\sc evtgen} \cite{Lange2001,Ping2008} with branching fractions set to the world average values~\cite{PDG}, and by {\sc lundcharm}~\cite{JCChen2000} for the remaining unknown decays. Each MC-generated event is mixed with a randomly triggered event recorded during data taking in order to  include effects of background contamination, such as beam-related background and cosmic rays, as well as electronic noise and hot wires. The analysis is performed in the framework of the BESIII offline software system which takes into account the detector calibration, event reconstruction and data storage.

Exclusive MC samples of $0.5$ million events each are generated for the processes $\psi(3686)\rightarrow\phi\eta^\prime, \phi f_1(1285), \phi\eta(1405)$, with $\phi\rightarrow K^+K^-$ and $\eta^\prime, f_1(1285), \eta(1405)\rightarrow\pi^+\pi^-\eta$, and $\eta\rightarrow\gamma\gamma$.  They are used in the optimization of the selection criteria and the determination of the detection efficiencies. The decays of $\psi(3686)\rightarrow\phi\eta^\prime, \phi\eta(1405)$ are generated using a helicity amplitude model \cite{Ping2008}. 
In the simulation of  $\psi(3686)\rightarrow\phi f_1(1285)$, the same model as for  $J/\psi\rightarrow\phi f_1(1285)$~\cite{Jpsiprd} is used.

\section{Data Analysis}
\subsection{\label{3.2}Event selection}
To select candidate events of the process $\psi(3686)\rightarrow\phi\pi^{+}\pi^{-}\eta$ with $\phi\rightarrow K^{+}K^{-}$ and $\eta\rightarrow\gamma\gamma$, the following criteria are imposed on the data and MC samples.
We select charged tracks in the MDC within the polar angle range $|\text{cos}\theta|<0.93$ and require that the points of closest approach to the beam line be within $\pm 10$ cm of the interaction point in the beam direction and within 1.0 cm in the plane perpendicular to the beam. The TOF and the specific energy loss, $dE/dx$, of a particle measured in the MDC are combined to calculate particle identification (PID) probabilities for pion, kaon and proton hypotheses. The particle type with the highest probability is assigned to each track. In this analysis, two kaon and two pion tracks with opposite charges are required.

Photon candidates are reconstructed from isolated clusters of energy deposits in the EMC. The energy deposited in nearby TOF counters is included to improve the photon reconstruction efficiency and energy resolution. At least two photon candidates are required, with a minimum energy of 25 MeV for barrel showers ($|\text{cos}\theta|<0.80$) and 50 MeV for end-cap showers ($0.86<|\text{cos}\theta|<0.92$). To exclude showers due to  charged particles, the angle between the nearest charged track and the shower in EMC must be greater than $10^\circ$. An EMC shower timing requirement, $0\leq t\leq 700$ ns,  is applied to suppress electronic noise and energy depositions unrelated to the event.

A four-constraint (4C) kinematic fit using four-momentum conservation is performed under the $\psi(3686)\rightarrow K^{+}K^{-}\pi^{+}\pi^{-}\gamma\gamma$ hypothesis. In events with more than two photon candidates, all pairs are tried and the combination with the smallest $\chi^2_{4\text{C}}$ value is retained. An event is rejected if $\chi^{2}_{4\text{C}} > 80$.

\begin{figure*}
\subfigure{
 \label{scatter}
 \includegraphics[width=2.0in]{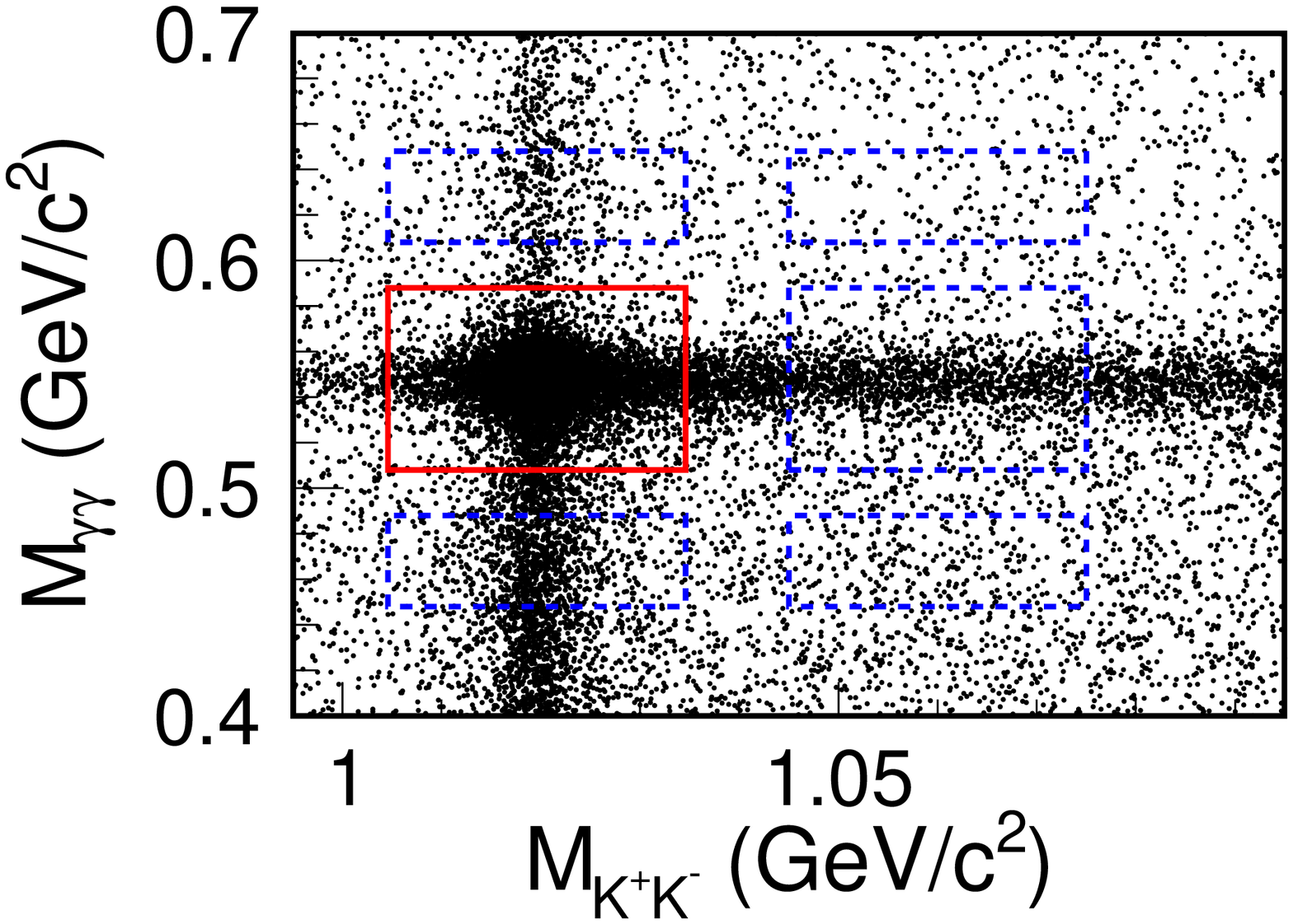}
 \put(-30,85){(a)}
 }
 \subfigure{
  \label{etacut}
 \includegraphics[width=2.0in]{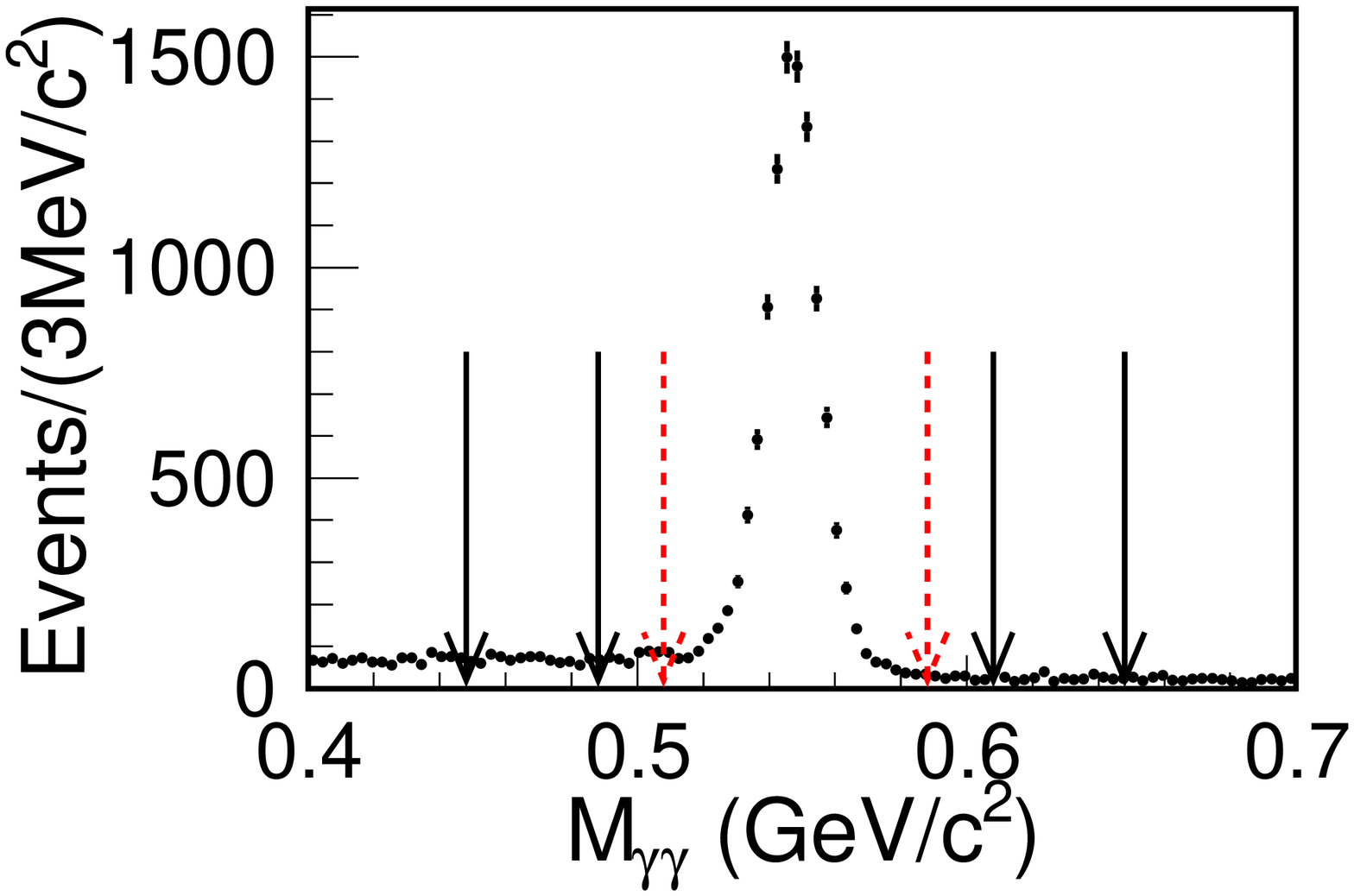}
  \put(-30,85){(b)}
 }
 \subfigure{
 \label{phicut}
 \includegraphics[width=2.0in]{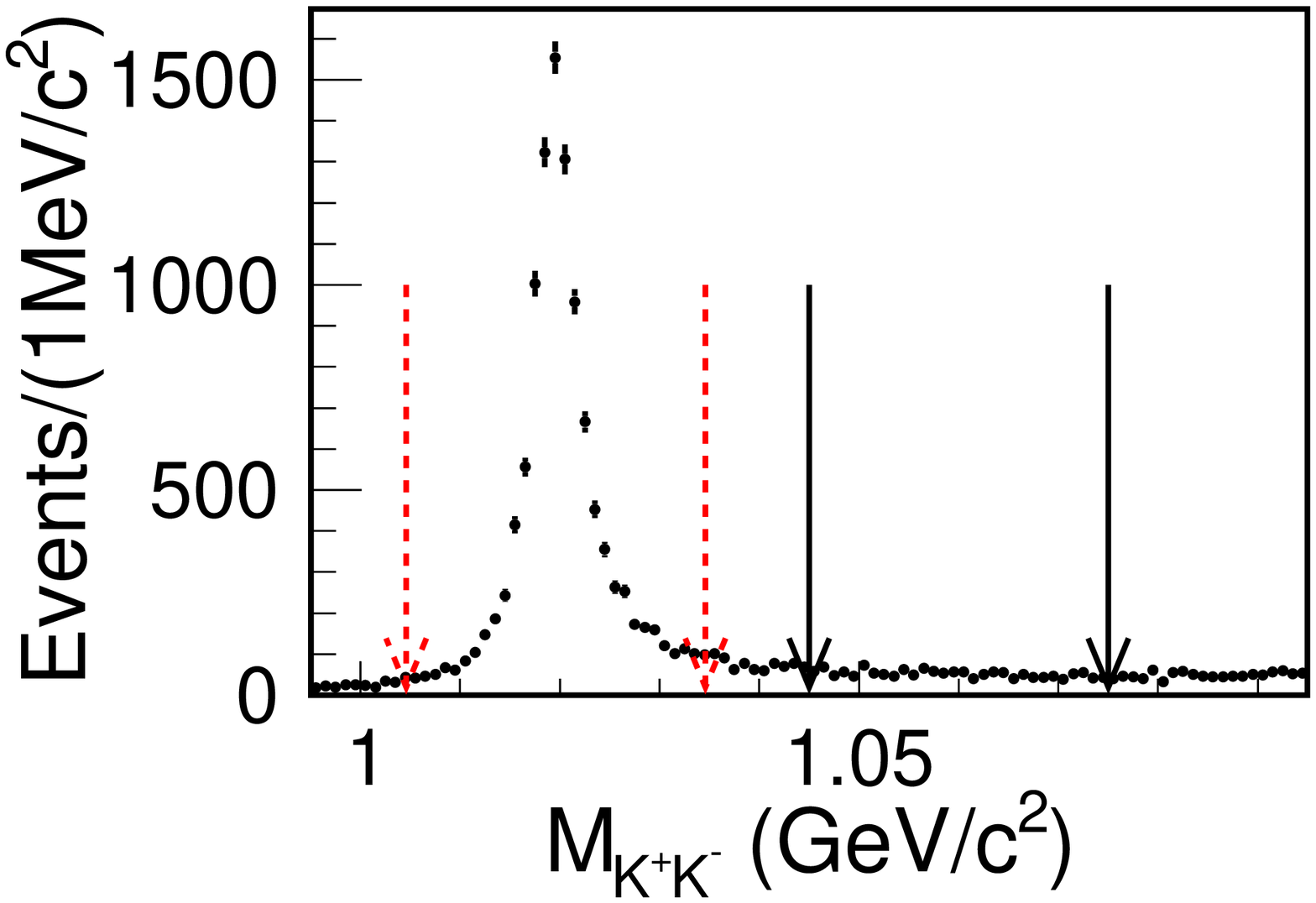}
 \put(-30,85){(c)}}
 \caption{(a) Distribution of $M_{\gamma\gamma}$ versus $M_{K^{+}K^{-}}$, where the red solid box shows the signal region and the blue dotted boxes are for  the sideband regions of $\eta$ and $\phi$.  (b) Distribution of $M_{\gamma\gamma}$ within the $\phi$ signal region. (c) Distribution of $M_{K^+K^-}$ within the $\eta$ signal region. The dashed arrows show the signal regions and solid arrows show the sideband regions, as described in the text.}\label{1}
\end{figure*}

The resulting distribution of the invariant mass $M_{\gamma\gamma}$ versus $M_{K^+K^-}$ is illustrated  in Fig.~\ref{scatter}; the area indicated by the solid box corresponds to the $\psi(3686)\rightarrow\phi\pi^+\pi^-\eta$ signal region. The distributions of $M_{\gamma\gamma}$  and $M_{K^+K^-}$ are shown in Figs.~\ref{etacut} and \ref{phicut}, respectively, where the $\eta$ and $\phi$ peaks are clearly observed.
The $\phi$ and $\eta$ signal regions are defined as $|M_{K^+K^-}-m_\phi|<0.015$ and $|M_{\gamma\gamma}-m_\eta|<0.040 \ \text{GeV}/c^2$, where $m_\phi$ and $m_\eta$ are the world average values of the $\phi$ and $\eta$ masses~\cite{PDG}.

The $\pi^{+}\pi^{-}\eta$ invariant mass distribution in the $\eta^\prime$ region is shown in Fig.~\ref{sumInMC_etap}. The main background contribution to the $\pi^{+}\pi^{-}\eta$ invariant mass region above 1.1 GeV/$c^2$ comes from the $\psi(3686)\rightarrow\pi^+\pi^-J/\psi$, $J/\psi\rightarrow K^+K^-\gamma\gamma$, and $\psi(3686)\rightarrow \gamma\gamma J/\psi$, $J/\psi\rightarrow K^+K^-\pi^+\pi^-$. To suppress this background, we require that both $M_{K^+K^-\gamma\gamma}$ and $M_{K^+K^-\pi^+\pi^-}$ are not in the $J/\psi$ mass region of  $[3.047,3.147]$ GeV/$c^2$ when $M_{\pi^+\pi^-\eta}>1.1$ GeV/$c^2$. The resulting $\pi^{+}\pi^{-}\eta$ invariant mass distribution is shown in Fig.~\ref{sumInMC}, where significant $f_1(1285)$ and $\eta(1405)$ peaks are observed.

\subsection{Background study}

\begin{figure*}
\centering
\subfigure{
 \label{sumInMC_etap}
 \includegraphics[width=2.5in]{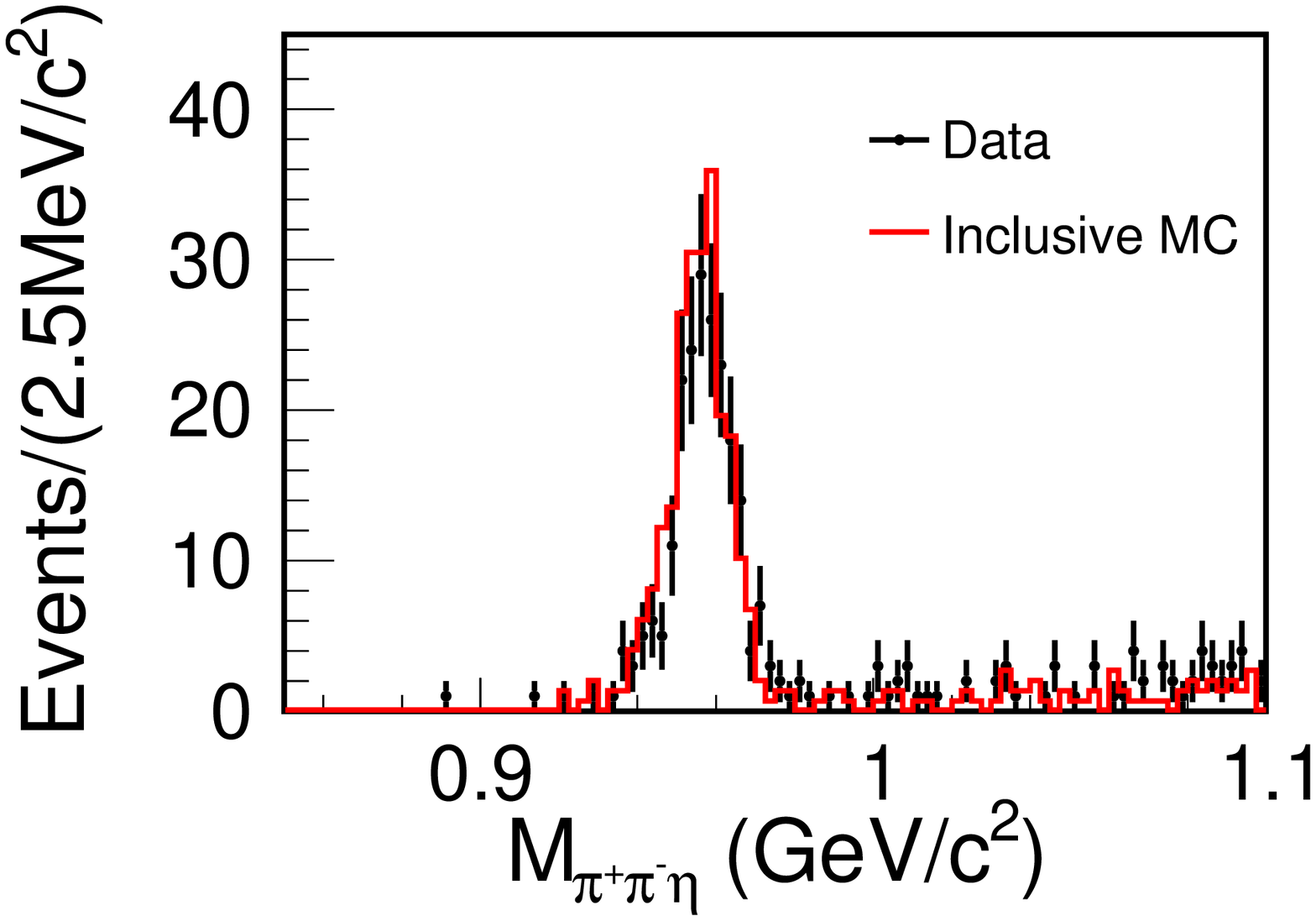}
 \put(-120,100){(a)}}
 \subfigure{
 \label{sumInMC}
 \includegraphics[width=2.5in]{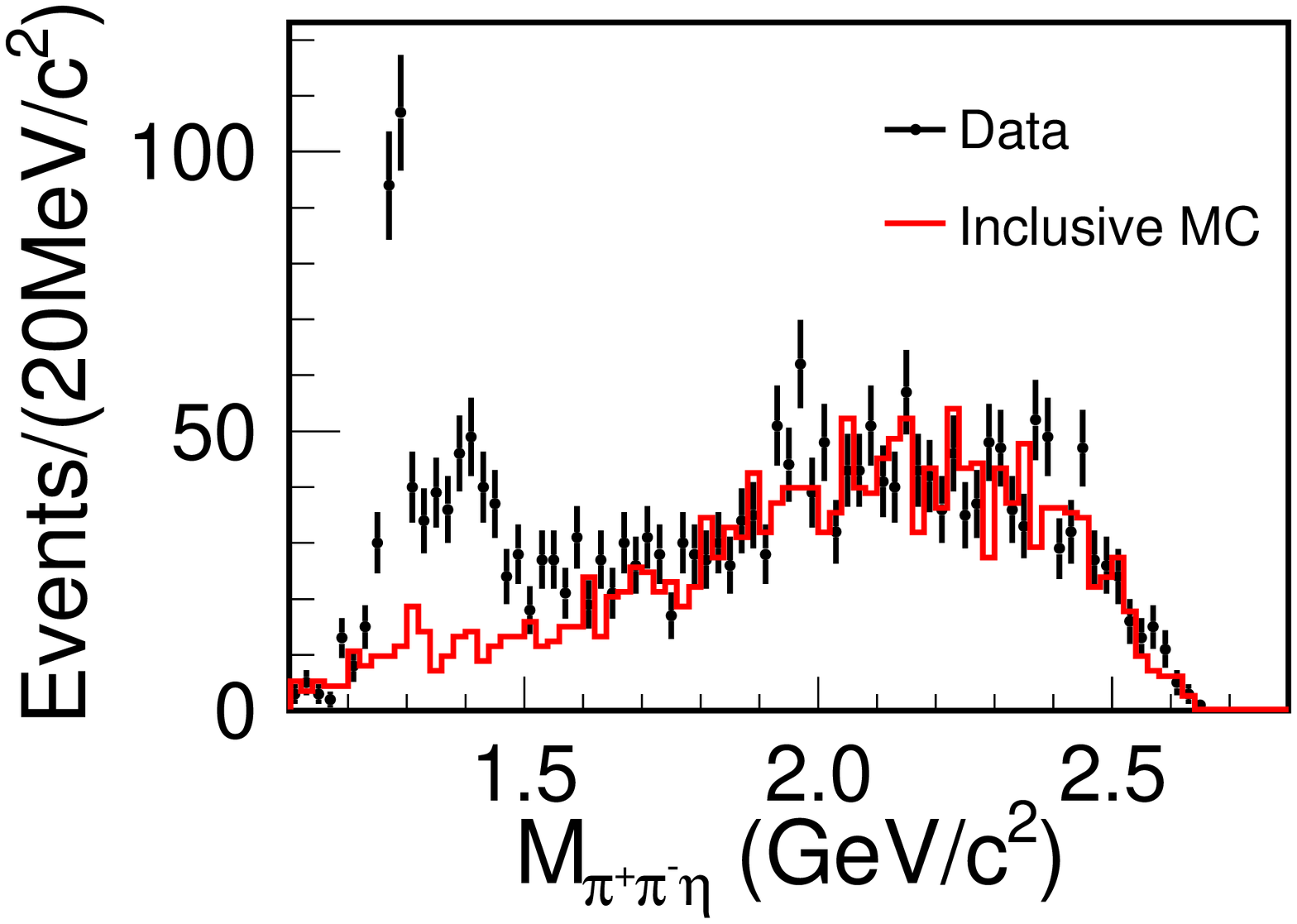}
 \put(-120,100){\footnotesize(b)}}
 \caption{Distribution of $M_{\pi^+\pi^-\eta}$ in $[0.85, 1.10]$ GeV/$c^2$ and $[1.1, 2.8]$ GeV/$c^2$ regions. The dots with error bars show data. The histogram shows the inclusive MC, scaled to the total number of events in data.}\label{InMC}
\end{figure*}

To investigate the background events, we apply the same selection to the inclusive MC sample of $506 \times10^6$ $\psi(3686)$ events. The $\pi^+\pi^-\eta$ invariant mass distribution of the selected events is displayed in Fig.~\ref{InMC}. The BF of $\psi(3686)\rightarrow\phi\eta^\prime$ in the MC was adjusted to provide agreement in the number of events with the data. For the mass region of $M_{\pi^+\pi^-\eta}> 1.1$ GeV/c$^2$, the contribution from the inclusive MC events is reasonably smooth, which indicates that the $f_1(1285)$ and $\eta(1405)$ peaks observed in data are not from the known decays of $\psi(3686)$ decays listed by the Particle Data Group (PDG)~\cite{PDG}.

To further study the background events, we estimate them with the $\eta$ - $\phi$ two-dimensional sideband. The $\eta$ sideband is defined by $0.448$ $\text{GeV}/c^2<M_{\gamma\gamma}<0.488$ $\text{GeV}/c^2$ or $0.608$ $\text{GeV}/c^2<M_{\gamma\gamma}<0.648$ $\text{GeV}/c^2$, and the $\phi$ sideband is defined by $1.045$ $\text{GeV}/c^2<M_{K^+K^-}<1.075$ $\text{GeV}/c^2$, as indicated by the dashed boxes in Fig.~\ref{scatter}.  There are no peaks evident for $\eta^\prime$, $f_1(1285)$ and $\eta(1405)$ from the sidebands.

For the background events from the  continuum process $e^+e^-\rightarrow\phi\pi^+\pi^-\eta$, we perform a study with the sample of $(2.93\pm0.01)$ fb$^{-1}$~\cite{psi3770data} taken at $\sqrt{s}=3.773$ GeV. After the same event selection as described above, clear $\eta^\prime$, $f_1(1285)$ and $\eta(1405)$ peaks are seen in the $\pi^+\pi^-\eta$ mass spectrum recoiling against the $\phi$. An unbinned maximum likelihood fit, analogous to the one in Sec.~\ref{measurement}, yields $221\pm15$, $26.8\pm7.1$, and $87\pm16$ events for $\eta^\prime$, $f_1(1285)$ and $\eta(1405)$, respectively. We assume that the observed signals come directly from $e^+e^-$ annihilations and not $\psi(3770)$ decays. A scale factor $f$ is defined as the ratio of the observed number $N$ in $\psi(3686)$ data to that in the $\psi(3770)$ data,
\begin{equation*}
 f\equiv\frac{N_{\psi(3686)}}{N_{\psi(3770)}}=\frac{\mathcal{L}_{\psi(3686)}}{\mathcal{L}_{\psi(3770)}}\cdot\frac{\sigma_{\psi(3686)}}{\sigma_{\psi(3770)}}\cdot\frac{\varepsilon_{\psi(3686)}}{\varepsilon_{\psi(3770)}},
\end{equation*}
where $N$, $\mathcal{L}$, $\sigma$ and $\varepsilon$ refer to the observed number events, integrated luminosity of data
samples, cross section and detection efficiency at the two c.m. energies. The details on the cross section can be found in Ref.~\cite{cross}. The detection efficiency ratios $\frac{\varepsilon_{\psi(3686)}}{\varepsilon_{\psi(3770)}}$ can be determined by Monte Carlo simulations. The scale factors are calculated to be 0.232, 0.242, and 0.236 and the normalized numbers of continuum events for $e^+e^-\rightarrow\phi\eta^\prime$, $e^+e^-\rightarrow\phi f_1(1285)$ and $e^+e^-\rightarrow\phi\eta(1405)$ at 3.686 GeV are determined to be $51.3\pm 3.5$, $6.5\pm 1.7$, and $20.5\pm3.8$, respectively. 
Due to identical event topology, these background events are indistinguishable from signal events and are subtracted directly from our nominal yields.
\subsection{\label{measurement}Measurement of BFs of \boldmath$\psi(3686)\rightarrow\phi\eta^\prime, \phi f_1(1285) {\rm ~and~} \phi\eta(1405)$}
\begin{figure}
\centering
\subfigure{
 \label{fitetap3686}
 \includegraphics[width=2.5in]{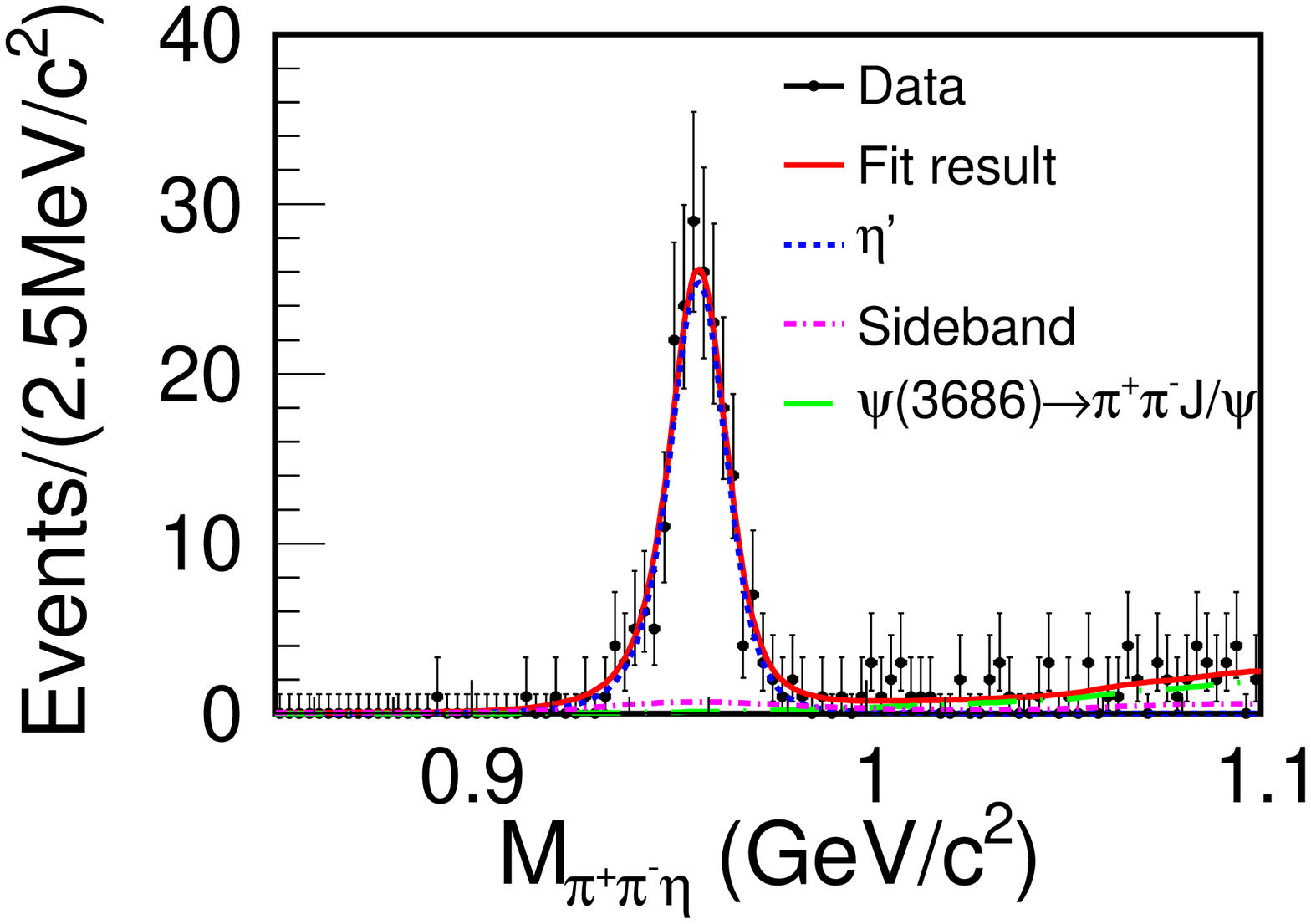}
 \put(-120,100){(a)}}
 \subfigure{
 \label{fitf1andeta14053686}
 \includegraphics[width=2.5in]{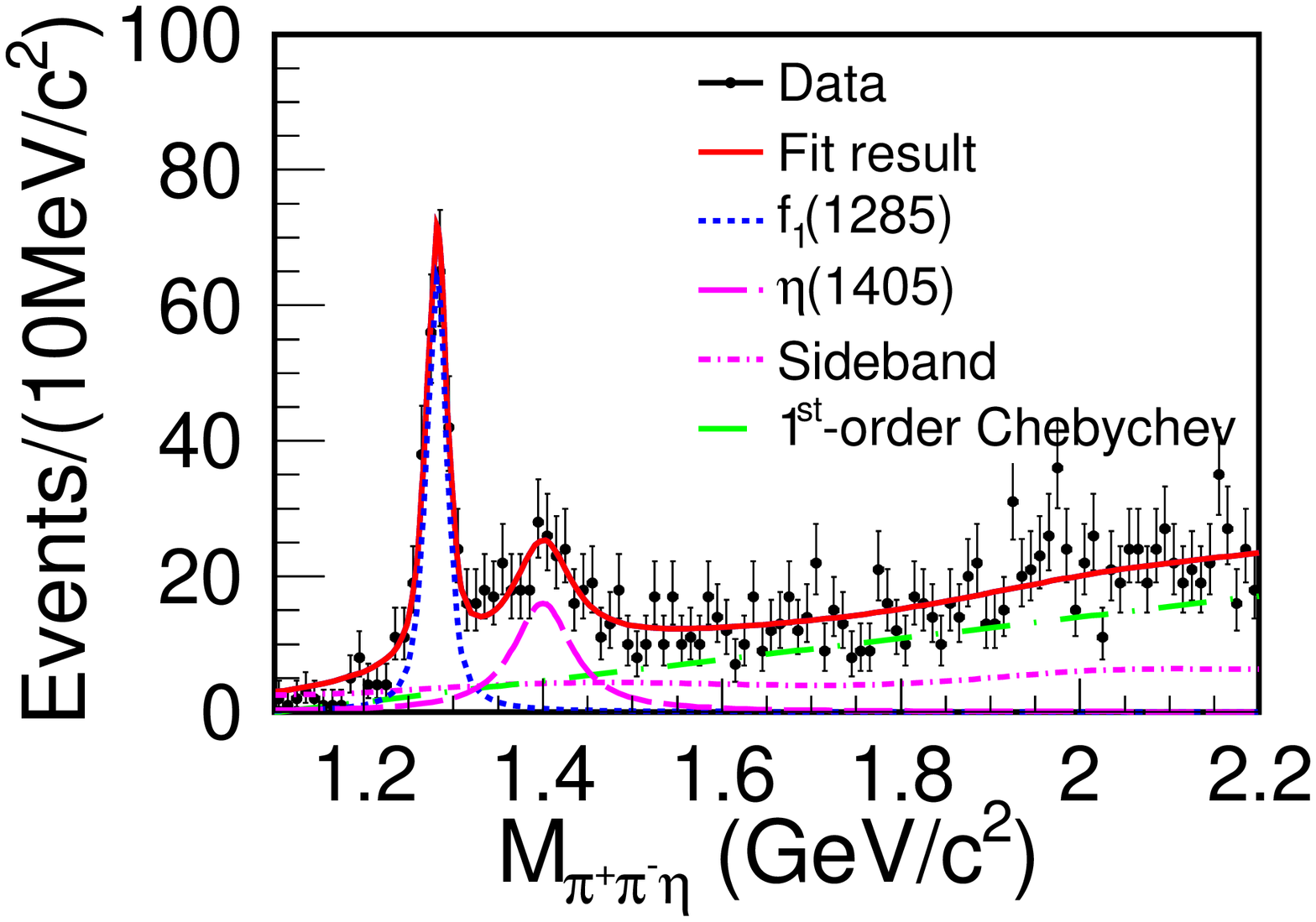}
 \put(-120,100){\footnotesize(b)}}
 \caption{Results  of the fits to $M_{\pi^+\pi^-\eta} $ in the ranges  of $[0.85, 1.10]$ GeV/c$^2$ (a) and $[1.1, 2.2]$ GeV/c$^2$ (b) , where the dots with error bars are data, and the curves are the results of the fit described in the text.}\label{fitresults}
\end{figure}

Since the $\eta^\prime$ signal is well isolated from the $f_1(1285)$ and $\eta(1405)$ peaks, we perform an extended unbinned maximum likelihood fit  to the $\pi^+\pi^-\eta$ invariant mass in the range of $[0.85, 1.10]$ GeV/$c^2$ to obtain the signal yields of $\eta^\prime$. In the fit, the total probability density function consists of a signal and a background contribution. The signal component is modeled from the MC-simulated signal shape using a nonparametric method \cite{Cranmer:2000du}, convolved with a Gaussian function to account for different mass resolutions in data and MC simulation. The background contribution is described by the two-dimensional $\eta$ - $\phi$ sideband and the background from $\psi(3686)\rightarrow\pi^+\pi^-J/\psi$.  The fit, shown in Fig.~\ref{fitetap3686}, yields $201\pm15$ $\phi\eta^\prime$ events.

Another fit to the $\pi^{+}\pi^{-}\eta$ invariant mass in the range of $[1.1, 2.2]$ GeV/$c^{2}$ is performed to obtain the signal yields of  $f_1(1285)$ and $\eta(1405)$ with an assumption of no interference between them.  In this case,  the total probability density function can be described with
 \begin{widetext}
 \begin{equation}
 \varepsilon(M_{\pi\pi\eta})\times[BW(M_{1},
 \Gamma_{1})\otimes G(\sigma_{1}) + BW(M_{2},\Gamma_{2})\otimes G(\sigma_{2})]  +BKG,
\end{equation}
\end{widetext}
where $BW(M, \Gamma) = \frac{\Gamma^2/4}{(M_{\pi\pi\eta}-M)^2+\Gamma^2/4}$ is the Breit-Wigner function representing the $f_1(1285)$ and $\eta(1405)$ signal shape, and $M_{1}$ [$M_{2}$] and $\Gamma_{1}$ [$\Gamma_{2}$] are the  mass and width for $f_{1}(1285)$ [$\eta(1405)$], which are free parameters in the fit. The Gaussian function $G$ represents the mass resolution, and the corresponding parameters, $\sigma_1$ and $\sigma_2$, are taken from the  MC simulations. The detection efficiency $\varepsilon(M_{\pi\pi\eta})$ as a function of the $\pi^{+}\pi^{-}\eta$ invariant mass is obtained from the MC simulation. $BKG$ refers to the following background components: 1) a smoothed shape from the two-dimensional $\eta$ - $\phi$ sidebands with fixed normalization, and 2) a linear polynomial describing the remaining background events.

The fit, shown in Fig.~\ref{fitf1andeta14053686}, yields $234\pm22$ $\phi f_1(1285)$ events and  $195\pm28$ $\phi \eta(1405)$ events. The corresponding statistical significances for $f_1(1285)$ and $\eta(1405)$ are $18\sigma$ and $9.7\sigma$, respectively. They are determined from differences of the likelihood values and the degrees of freedom between the fits with and without the resonance.
The  mass  and width of the $f_1(1285)$ [$\eta(1405)$] determined in the fit are $1289.3\pm2.8$ $[1404.0\pm11.0]$ MeV/c$^{2}$ and  $17.1\pm3.4$ [$79.0\pm16.0 $] MeV, respectively. They are in reasonable agreement with the world average values~\cite{PDG}, but with larger uncertainties due to our limited statistics.  

The signal yields, detection efficiencies obtained from the MC simulations, the background contribution from the continuum process, and the BFs of  $\psi(3686)\rightarrow\phi\eta^\prime$, $\psi(3686)\rightarrow\phi f_1(1285)$ and $\psi(3686)\rightarrow\phi\eta(1405)$ are listed in Table~\ref{tableBF}.

To test the ``$12\%$ rule'', the ratios $Q_h=\frac{\mathcal{B}(\psi(3686)\rightarrow h)}{\mathcal{B}(J/\psi\rightarrow h)}$ are determined to be $(3.28\pm 0.56)\%$, $(11.32\pm2.66)\%$ and $(4.23\pm2.27)\%$ for each channel, which are summarized in Table~\ref{tableBF}. In the calculation of the ratio, the $J/\psi$ branching fractions are taken from the PDG~\cite{PDG}.

\begin{table*}
\renewcommand{\arraystretch}{1.2}
\centering
 \caption{Summary of signal yields, estimated background events from the continuum, statistical significances, detection efficiencies, corresponding branching fraction and the values of $Q_h=\frac{\mathcal{B}(\psi(3686)\rightarrow h)}{\mathcal{B}(J/\psi\rightarrow h)}$}.
 \label{tableBF}
\noindent
 \begin{tabular}{p{3.5cm}<{\centering}p{1.5cm}<{\centering}p{1.5cm}<{\centering}p{2.0cm}<{\centering}p{2.0cm}<{\centering}p{3.5cm}<{\centering}p{3.5cm}<{\centering}}
 \hline\hline
Decay mode  & $N_\text{obs}$ & $N_\text{continuum}$ & Significance & Efficiency(\%) & Branching fraction &$Q_h$ \\
 \hline
$\psi(3686)\rightarrow\phi\eta^\prime$                                       & $201\pm15$ & $51.3\pm3.5$ &$\cdots$ &$26.8\%$ & $(1.51\pm0.16\pm0.12)\times10^{-5}$ & $(3.28\pm 0.56)\%$\\
\hline
$\psi(3686)\rightarrow\phi f_1(1285),$ $f_1(1285)\rightarrow\pi^+\pi^-\eta$  & \multirow{2}{*}{$234\pm22$} & \multirow{2}{*}{$6.5\pm1.7$} & \multirow{2}{*}{$18\sigma$} & \multirow{2}{*}{$25.6\%$} & \multirow{2}{*}{$(1.03\pm0.10\pm0.09)\times10^{-5}$}& \multirow{2}{*}{$(11.32\pm 2.66)\%$}\\
\hline
$\psi(3686)\rightarrow\phi\eta(1405),$ $\eta(1405)\rightarrow\pi^+\pi^-\eta$ & \multirow{2}{*}{$195\pm28$} & \multirow{2}{*}{$20.5\pm3.8$} & \multirow{2}{*}{$9.7\sigma$} & \multirow{2}{*}{$23.9\%$} & \multirow{2}{*}{$(8.46\pm1.37\pm0.92)\times10^{-6}$}& \multirow{2}{*}{$(4.23\pm 2.27)\%$}\\
 \hline\hline
\end{tabular}
\end{table*}

\section{systematic uncertainties}
The sources of systematic uncertainties include the efficiency difference between data and MC simulation for charged track reconstruction, photon detection, PID requirements and kinematic fit as well as input branching fractions, the number of $\psi(3686)$ events, and yield fitting procedures. The corresponding contributions to the measurement of the branching fractions are discussed in detail below.

\begin{table}
\centering
 \caption{Summary of sources of systematic uncertainties and their corresponding contributions in \%.}
 \label{tablesys}
 \vspace{0.2cm}
\noindent
\footnotesize
 \begin{tabular}{lccc}
 \hline\hline
 Sources  & $\eta^{\prime}$ & $f_{1}(1285)$ & $\eta(1405)$ \\
 \hline
 Charged tracks & 4.0 & 4.0 & 4.0 \\

 Photon detection & 2.0 & 2.0 & 2.0\\

 PID  & 4.0 & 4.0 & 4.0 \\

 Kinematic fit & 0.3 & 0.2& 0.2 \\

 $\mathcal{B}(\phi\rightarrow K^{+}K^{-})$&1.0&1.0&1.0 \\

 $\mathcal{B}(\eta\rightarrow\gamma\gamma)$&0.5&0.5&0.5 \\

 $\mathcal{B}(X\rightarrow\pi^{+}\pi^{-}\eta)$&1.6&$\cdots$&$\cdots$\\

 Number of  $\psi(3686)$ events &0.6&0.6&0.6\\

 $\phi$ mass window&1.2 &1.2&1.2\\

 $\eta$ mass window & 0.3&0.3 &0.3\\

 $J/\psi$ veto    & $\cdots$ & 2.7 & 3.2\\

 Fit range & 2.7 &1.3 &6.9\\

 Signal shape &2.7 &1.8 &1.7\\

 Background polynomial &$\cdots$ &0.0&4.2\\

 Sideband &1.7 &1.7&1.2\\

 Parameters of $\phi f_1(1285)$ generation &$\cdots$ &4.7 &$\cdots$\\
\hline
 Total &7.7 &8.7 &10.9\\
 \hline\hline
\end{tabular}
\end{table}

\begin{enumerate}[(a)]
	\item \textit{MDC tracking efficiency:} The charged tracking efficiency has been investigated with the clean control channels $J/\psi\rightarrow\pi^{+}\pi^{-}p\bar{p}$ and $J/\psi\rightarrow\rho\pi$ \cite{PIDefficicy}. It is found that the MC simulation agrees with data within 1\% for each charged track. Therefore, 4\% is taken as the systematic uncertainty from the four charged tracks in the final state.
	
	\item \textit{Photon detection efficiency:} The photon detection efficiency has been studied using a control sample of $J/\psi\rightarrow\rho\pi$~\cite{PIDefficicy}. The results indicate that the difference between the detection efficiencies of data and MC is around 1\% per photon. Thus, 2\% is taken as the total systematic uncertainty for the detection of the two photons in this analysis.
	
	\item \textit{PID efficiency:} To evaluate the PID efficiency uncertainty, we have studied the kaon and pion PID efficiencies using the clean control samples of $J/\psi\rightarrow K^{*\pm}K^{\mp}$ and $J/\psi\rightarrow\rho\pi$~\cite{PIDefficicy}, respectively. We find that the difference in the PID efficiency between data and MC is 1\% for each kaon or pion. Hence, 4\% is taken as the total systematic uncertainty from the PID efficiency.
	
	\item \textit{Kinematic fit:}  The uncertainty associated with the 4C kinematic fit comes from the inconsistency between data and MC simulation of the fit; this difference is reduced by correcting the track helix parameters of the MC simulation. Following the method described in Ref.~\cite{Ablikim2013_1}, we obtain the systematic uncertainties for the 4C kinematic fit as 0.3\%, 0.2\% and 0.2\% for the branching fractions of $\psi(3686)\rightarrow\phi\eta^\prime$, $\psi(3686)\rightarrow\phi f_1(1285)$ and $\psi(3686)\rightarrow\phi\eta(1405)$, respectively.

	\item \textit{Intermediate decay branching fractions:} The branching fractions of $\phi\rightarrow K^{+}K^{-}, \eta\rightarrow\gamma\gamma$, and $\eta^{\prime}\rightarrow\pi^{+}\pi^{-}\eta$ are taken from the PDG~\cite{PDG}. The uncertainties of these branching fractions, 1.0\%, 0.5\%, and 1.6\%, respectively, are taken as the systematic uncertainties.
	
	\item \textit{Number of $\psi(3686)$ events:} The number of $\psi(3686)$ events is determined from an analysis of inclusive hadronic $\psi(3686)$ decays. The uncertainty of the number of $\psi(3686)$ events, 0.6\% \cite{psipdata}, is taken as the systematic uncertainty in the calculation of the BFs.
	
	\item \textit{$\phi$ mass window:} In Ref.~\cite{Ablikim2016}, a control
sample of $J/\psi \rightarrow \phi \eta^\prime, \phi \rightarrow K^+K^-, \eta^\prime \rightarrow \gamma \pi^+\pi^-$ is used to study the uncertainty due to the $\phi$ mass window requirement.
We adopt the resulting uncertainty of 1.2\% from that study.
	
	\item \textit{$\eta$ mass window:}  To estimate the uncertainty from the $\eta$ mass requirement, we select a clean sample of $J/\psi\rightarrow\phi\eta$ without this requirement. Events with two oppositely charged tracks and two good photons are selected. The charged tracks must be identified as kaons. A 4C kinematic fit is performed with the $J/\psi\rightarrow K^+K^-\gamma\gamma$ hypothesis and the $\chi^{2}_{4\text{C}}$ is required to be less than 40. The $K^+K^-$ invariant mass is required to be in the $\phi$ mass region, $|M_{K^{+}K^{-}}-m_\phi|<0.015$ GeV/$c^2$. We perform a fit to the mass spectrum of $\gamma\gamma$, where a Crystal Ball function~\cite{CB} is used to describe the $\eta$ signal and a first-order Chebyshev polynomial describes the background. Requiring $|M_{\gamma\gamma}-m_{\eta}|<0.04$ GeV/c$^2$, we regard the difference of the $\eta$ selection efficiencies between data and MC samples, 0.3\%,  as the systematic uncertainty.
	
	\item \textit{$J/\psi$ veto:} To remove the background events from $\psi(3686)\rightarrow\pi^+\pi^-J/\psi$, we have applied a requirement of $|M_{K^+K^-\gamma\gamma}-M_{J/\psi}|>0.05$ GeV/$c^2$. 	
	In order to estimate the systematic uncertainty, this requirement is varied by $\pm$ 0.01 GeV/c$^2$ for both $\psi(3686)$ and continuum data. The maximum changes to the nominal results, 2.7\% and 3.2\%, respectively, for $f_1(1285)$ and $\eta(1405)$ are taken as the systematic uncertainties.
	
	\item \textit{Fit range:} We perform alternative fits for $\psi(3686)$ and continuum data by varying the fit ranges, and assign the maximum change of the results, 2.7\%, 1.3\% and 6.9\%, as the systematic uncertainties.
	
	\item \textit{Signal shape:}  To obtain the number of $\phi\eta^{\prime}$ events in the fit to $M_{\pi^+\pi^-\eta}$, the MC shape of the $\eta^\prime$ convolved with a Gaussian function is used to describe the signal shape. In order to estimate the systematic uncertainty due to this shape, alternative fits are performed to determine the yields of signal and peaking background events, replacing the MC shape with a Breit-Wigner function. The change of the result, 2.7\%, is taken as the systematic uncertainty. The uncertainties from the signal shape of the $f_1(1285)$ and $\eta(1405)$ are estimated by varying the mass resolutions by $\pm$10\%, in both $\psi(3686)$ and continuum data, to account for the difference between data and MC simulation. We take the changes of the signal yields of $\phi f_1(1285)$ and $\phi\eta(1405)$ events, 1.8\% and 1.7\%, as the systematic uncertainties.

	\item \textit{Background shape from Chebyshev polynomial:} To estimate the uncertainty of background shape in the fit to $M_{\pi^+\pi^-\eta}$,  we performed alternative fits by replacing the first-order Chebyshev polynomial with a second-order Chebyshev polynomial for both $\psi(3686)$ and continuum data.  The changes of 0.0\% and 4.2\% are taken as systematic uncertainties.

	\item \textit{Sideband:} The uncertainty on the $\phi \eta^\prime$ yield caused by the sideband regions is estimated by changing those regions to $0.447$ $<M_{\gamma\gamma}<0.487$, $0.609$ $<M_{\gamma\gamma}<0.649$, and $1.046$ $<M_{K^+K^-}<1.076$ GeV/$c^2$. The change of the yields, 1.7\%, is regarded as the systematic uncertainty for that mode. For the determination of the  $f_1(1285)$ and $\eta(1405)$ signal yields, the background events estimated from the two-dimensional $\eta$ - $\phi$ sidebands have only a smooth contribution under the  $f_1(1285)$ and $\eta(1405)$ peaks. To estimate the uncertainty associated with the sidebands, we performed an alternative fit by removing the constraint on the number of the background events estimated from the two-dimensional $\eta$ - $\phi$ sidebands. The changes to the nominal results, 1.7\% and 1.2\% for $f_1(1285)$ and $\eta(1405)$, respectively, are considered as systematic uncertainties.

	\item\textit{Parameters of $\phi f_1(1285)$ generation:} Due to the limited statistics, and because $\psi(3686)\rightarrow\phi f_1(1285)$ is expected to be similar to the process of $J/\psi\rightarrow\phi f_1(1285)$, we use the same model for $J/\psi\rightarrow\phi f_1(1285)$~\cite{Jpsiprd} to generate the signal MC sample of $\psi(3686)\rightarrow\phi f_1(1285)$ to determine the detection efficiency, and the angular distribution of this decay can be expressed as $\frac{dN}{d\text{cos}\theta}=1+\alpha \text{cos}^2\theta$. The parameter $\alpha$ used in the MC sample generation is taken from the angular distribution of $\phi$ in the rest frame of $J/\psi$ found in real data. Following the method described in Ref.~\cite{Jpsiprd}, the impact of the uncertainty of these parameters on the efficiency, 4.7\%, is taken as a source of systematic uncertainty on the branching fraction.
\end{enumerate}

A summary of the systematic errors is shown in Table~\ref{tablesys}. By assuming that all of them are independent, the total systematic uncertainty is obtained by adding the individual contributions in quadrature.

\section{summary}
Based on a sample of $448.1\times10^6$ $\psi(3686)$ events collected with the BESIII detector, we presented a study of  $\psi(3686)\rightarrow\phi\pi^{+}\pi^{-}\eta$.  The branching fraction of $\psi(3686)\rightarrow\phi\eta^\prime$ was determined to be $(1.51\pm0.16\pm0.12)\times 10^{-5}$, which is consistent with the previous measurement~\cite{phietap}, and the precision is significantly improved.

In addition, the $f_1(1285)$ and $\eta(1405)$ are also clearly observed in the $\pi^+\pi^-\eta$ mass spectrum with statistical significances of $18\sigma$ and $9.7\sigma$. Using a fit assuming no interference between them, the resulting masses and widths of these resonances are in reasonable agreement with the world average values. The product branching fractions were measured for the first time to be $\mathcal{B}(\psi(3686)\rightarrow\phi f_{1}(1285),f_{1}(1285)\rightarrow\pi^{+}\pi^{-}\eta) =(1.03\pm0.10\pm0.09)\times 10^{-5}$ and $\mathcal{B}(\psi(3686)\rightarrow\phi\eta(1405),\eta(1405)\rightarrow\pi^{+}\pi^{-}\eta) =(8.46\pm1.37\pm0.92)\times 10^{-6}$. It is interesting that the $\eta(1405)$ is not significant in the $\pi^+\pi^-\eta$ mass spectrum recoiling against a $\phi$ in $J/\psi$ decays~\cite{Jpsiprd} but has a larger significance in $\psi(3686)$ decays. However, the low production rate of $\eta(1405)$ in $\psi(3686)\rightarrow\phi\pi^+\pi^-\eta$ still favors the conclusion, as reported in Ref.~\cite{Jpsiprd}, that $u$ and $d$ quarks account for more of the quark content in the $\eta(1405)$ than the $s$ quark.

From the results of the ratio $Q_h$ shown in Table~\ref{tableBF}, the decays of $\psi(3686)$ to $\phi\eta^\prime$ and $\phi\eta(1405)$ are suppressed by a factor of 3.6 and 2.8, respectively, compared with the ``$12\%$ rule'', while $\phi f_1(1285)$ is consistent with the rule. 

\begin{acknowledgements}
The BESIII collaboration thanks the staff of BEPCII and the IHEP computing center for their strong support. This work is supported in part by National Key Basic Research Program of China under Contract No. 2015CB856700; National Natural Science Foundation of China (NSFC) under Contracts Nos. 11335008, 11425524, 11625523, 11635010, 11675184, 11735014; the Chinese Academy of Sciences (CAS) Large-Scale Scientific Facility Program; the CAS Center for Excellence in Particle Physics (CCEPP); Joint Large-Scale Scientific Facility Funds of the NSFC and CAS under Contracts Nos. U1532257, U1532258, U1632107, U1732263; CAS Key Research Program of Frontier Sciences under Contracts Nos. QYZDJ-SSW-SLH003, QYZDJ-SSW-SLH040; 100 Talents Program of CAS; INPAC and Shanghai Key Laboratory for Particle Physics and Cosmology; German Research Foundation DFG under Contract No. Collaborative Research Center CRC 1044, FOR 2359; Istituto Nazionale di Fisica Nucleare, Italy; Koninklijke Nederlandse Akademie van Wetenschappen (KNAW) under Contract No. 530-4CDP03; Ministry of Development of Turkey under Contract No. DPT2006K-120470; National Science and Technology fund; The Knut and Alice Wallenberg Foundation (Sweden) under Contract No. 2016.0157; The Royal Society, UK under Contract No. DH160214; The Swedish Research Council; U. S. Department of Energy under Contracts Nos. DE-FG02-05ER41374, DE-SC-0010118, DE-SC-0012069; University of Groningen (RuG) and the Helmholtzzentrum fuer Schwerionenforschung GmbH (GSI), Darmstadt.
\end{acknowledgements}
\nocite{*}
\bibliographystyle{apsrev4-1}
\bibliography{draft}

\end{document}